\documentclass[11pt]{article}\usepackage{amsmath,amssymb,multirow,jheppub,graphicx}
\usepackage{centernot}
\usepackage{color}
\allowdisplaybreaks[1]
\usepackage{tikz}
 \tikzset{node distance=2cm, auto}
\usepackage{tikz-cd}

\renewcommand{\Re}{\text{Re }}
\renewcommand{\Im}{\text{Im }}

\def\Im{\text{Im}}
\def\Re{\text{Re}}

\def\tr{\text{tr}}

\def\bar{\overline}

\def\I{{\cal I}}

\def\R{{\mathbb R}}
\def\coeff#1#2{{\textstyle {\frac {#1}{#2}}}}

\def\half{\coeff 12}
\def\N{{\cal N}}
\usepackage{verbatim}   
\usepackage[all]{xy}
\usepackage{bm}  
\def\Dslash{{\rlap{\raise 1pt \hbox{$\>/$}}D}}
\def\Pslash{{\rlap{\raise  1pt \hbox{$\>/$}}\,\partial}}

\newcommand{\B}{{\cal B}}

\newcommand{\Ib}{{\overline\I}}

\renewcommand{\P}{{\mathbb P}}

\renewcommand{\Im}{\mathrm{Im}}
\renewcommand{\Re}{\mathrm{Re}}

\title{{New Methods in   QFT and QCD: From Large-N Orbifold Equivalence to  Bions and  Resurgence}}
\author[a]{Gerald  V. Dunne,} 
\author[b,c]{Mithat \"Unsal}
\affiliation[a]{Department of Physics, University of Connecticut, Storrs, 
CT 06269-3046, USA}
\affiliation[b]{Department of Physics, North Carolina State University, 
Raleigh, NC 27695, USA}
\affiliation[c]{Department of Mathematics,  Harvard University, Cambridge, MA, 02138, USA}
\abstract{
 We present a broad conceptual introduction to some new ideas  in non-perturbative QFT. 
The large-$N$ orbifold-orientifold equivalence connects a natural large-$N$ limit of QCD to QCD with adjoint fermions. 
QCD(adj) with periodic boundary conditions  
and double-trace deformation of Yang-Mills theory satisfy large-$N$ volume independence, a type of orbifold equivalence. 
Certain QFTs that satisfy volume independence at $N=\infty$  exhibit adiabatic continuity at   finite-$N$, 
and also become semi-classically calculable on small  $\mathbb R^3 \times S^1$. 
We discuss the role of  monopole-instantons, and magnetic and neutral bion saddles  in connection to  
 mass gap, and  center and chiral symmetry realizations. 
Neutral bions  also provide a weak coupling semiclassical realization of infrared-renormalons.  These considerations help motivate the necessity of complexification of path integrals (Picard-Lefschetz theory) in semi-classical analysis, and  highlights the importance of  hidden topological angles. Finally, we briefly review the  resurgence program, which potentially provides a novel non-perturbative continuum definition of QFT. All these ideas are continuously connected.   }


\begin{document}
\maketitle
%
%

%
%
%

\section{Introduction}

This review is a basic introduction  to some new  methods and ideas  in quantum gauge theories in four dimensions, and sigma models in two dimensions, with an underlying emphasis on making progress in the understanding of quantum chromodynamics (QCD).  The review has six short sections, each of which describes the most important concepts underlying these methods. The ideas are illustrated with simple examples.
In fact, all six stories are continuously connected, and  each borrows a new idea/perspective from the previous one.  Our intention is to provide a framework in quantum field theory (QFT) in which all reasonable methods,   lattice field theory, supersymmetric field theory, and  semi-classical continuum methods in  (non-supersymmetric) QFT,  among others, are treated on a similar footing, and can be used in conjunction to explain some of the mysteries of strongly interacting QFT.

\section{Large-$N$ Orbifold and Orientifold Equivalences}
\label{largeN}
\begin{figure}[ht]
\begin{center}
\includegraphics[angle=0, width=1.00\textwidth]{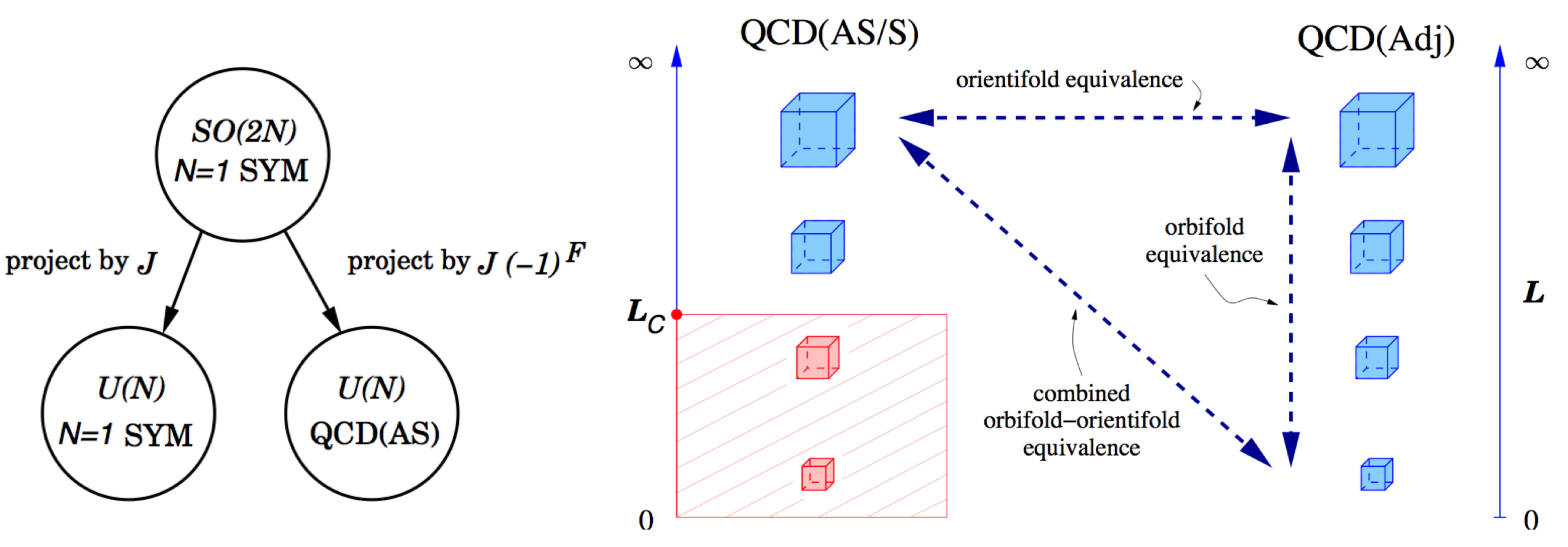}
\caption{
({\bf Left})  Large $N$   equivalences relating QCD(AS) and   ${\cal N}$=1 SYM. ({\bf Right}) 
Volume reduction and expansion are examples of orbifold projections, and center-symmetry realization governs the realization of the large-volume/small-volume equivalence (Eguchi-Kawai equivalence).
}
\label{orienti}
\end{center}
\end{figure}

Large-$N$ orbifold/orientifold equivalence is an exact equivalence between certain sub-sectors of seemingly unrelated quantum field  theories, in the limit of large-$N$, where $N$ is the number of internal degrees of freedom. $N$ can be either the rank of the color or global symmetry group.  These equivalences provide  new and important insights into the dynamics of strongly coupled theories. Interesting examples of equivalences are: 
(i) those relating a theory on an infinite lattice to a  theory on a
finite-size lattice, or even on a single-site lattice, (unitary matrix model) \cite{Eguchi:1982nm, GonzalezArroyo:1982hz,Kovtun:2007py}; 
or (ii) in the continuum, relating  a theory on $\R^d$ to a theory  compactified  on $\R^{d-k} \times (S^1)^k$; or (iii) equivalences relating a  supersymmetric theory to a non-supersymmetric one \cite{Kachru:1998ys, Bershadsky:1998cb, Armoni:2003gp,Armoni:2003fb,Armoni:2004ub,Kovtun:2003hr,Kovtun:2004bz,Unsal:2006pj,Kovtun:2005kh}. See 
Fig.\ref{orienti}.

The main idea is the following. 
One starts with some  ``parent" theory, and constructs ``daughter'' theories, by using  
certain orbifold/orientifold projections, which amount to retaining only those fields which are invariant under a particular discrete symmetry group of the parent theory. 
 For discrete abelian projections,  the planar graphs of the daughter theories  and  the parent theory coincide  
(up to a simple rescaling of the  coupling constant), which implies the coincidence of their
planar perturbative expansions at $N=\infty$ \cite{Kachru:1998ys,Bershadsky:1998cb}. 
 Clearly,  such a perturbative equivalence is an invitation to study possible non-perturbative realizations of this idea \cite{Armoni:2003gp,Armoni:2003fb}.  Refs. \cite{Kovtun:2003hr,Kovtun:2004bz} proved that the validity of the non-perturbative equivalence relies on certain symmetry realizations, as described below. 
 
To appreciate the implications of such equivalences, consider a simple example. 
Start with ${\cal N}=1$ Super-Yang-Mills (SYM) with  gauge group $SO(2N)$. The field content of the theory consists of a gauge boson, $A_{\mu}$, and a Weyl fermion, $\lambda$,  in the adjoint representation.  Two non-trivial ${\mathbb Z}_2$ projections of the    parent $SO(2N)$ theory 
lead to two distinct $U (N )$ gauge theories,  see Fig~\ref{orienti} \cite{Unsal:2006pj}.     
Let $J\equiv  i \sigma_2 \otimes 1_N \in SO(2N)$ denote the  symplectic form which is real and 
anti-symmetric. There are two natural 
 projections,   by $J$ and  $J(-1)^F$ where $(-1)^F$ is fermion number modulo two, 
 which may  be realized by the following constraints on the fields:
\begin{eqnarray}
J:  \;  & \; \;\;  A_\mu =JA_\mu J^T, \qquad  \lambda ={\color{red} +} J\lambda J^T, \qquad \Longrightarrow \quad U(N) \;\;   {\cal N}=1 \;  {\rm  SYM } 
\cr
J(-1)^F: \;  &A_\mu =JA_\mu J^T, \qquad  \lambda = {\color{red} -} J\lambda J^T, \qquad \Longrightarrow \quad U(N) \;\;   {\rm QCD(AS)} 
\end{eqnarray}
QCD(AS) is a  non-supersymmetric  gauge theory  with a single Dirac fermion  in the   two-index antisymmetric tensor representation  $\Psi_{ij}$. 
Clearly, the two daughter theories are different QFTs with different matter content, and different global symmetries. 
The parent and $J$-daughter theories are supersymmetric, while  the $J(-1)^F$-daughter theory is non-supersymmetric.   
In addition to the presence vs. absence of supersymmetry in the two daughter theories, the global discrete chiral symmetry and the center symmetries of these two daughter theories are also  different.

Before stating the main result, we note that the daughter theories also possess a non-trivial ${\mathbb Z}_2$ symmetry,  charge conjugation $C$, which is the image of the 
global part of the gauge symmetry in the parent QFT. (Think of embedding $U(N) \times U(N)^*$ into $SO(2N)$, and note that the $J$-action interchanges $U(N) \leftrightarrow U(N)^*$.)
This symmetry will be used to define a ${\mathbb Z}_2$  neutral sector  in the daughter theory, which constitutes the image of operators in the parent theory.  Clearly, ${\mathbb Z}_2$-odd operators are not in the image. In general orbifold projections, the discrete symmetry defining the neutral sector is the ${\mathbb Z}_k$ cyclic permutation symmetry. 
The operators transforming nontrivially under 
${\mathbb Z}_2$, or ${\mathbb Z}_k$, are also called `twisted' operators \cite{Kachru:1998ys,Bershadsky:1998cb}. 

\vspace{3mm}
\noindent
{\bf Necessary and sufficient conditions  \cite{Kovtun:2003hr,Kovtun:2004bz}:} There exists an exact  equivalence between the neutral sectors of parent-daughter pairs as well as daughter-daughter pairs {\it provided}:
\begin{itemize}
\item The symmetry in the parent theory used in the projection is  not spontaneously broken.  
\item   The symmetry  defining the neutral sector  in the daughter theory  
is not spontaneously broken. 
\end{itemize}
In the present example,  provided  the charge-conjugation $C$ is unbroken in QCD(AS), and the fermion number modulo two, $(-1)^F$,  is unbroken in $\N=1$, 
the $C$-even subsector of QCD(AS) is exactly equivalent to the bosonic sub-sector of  $\N=1$ SYM \cite{Unsal:2006pj}. The  $C$-realization is studied on the lattice 
\cite{DeGrand:2006qb},  demonstrating the existence of both $C$-broken and unbroken phases.

\vspace{3mm}
\noindent
{\bf Why is this useful?} 
 QCD(AS) is a natural generalization of $SU(3)$ QCD  to a large-$N$ limit \cite{Armoni:2003gp} , due to the simple fact that for SU(3), the fundamental (F)  and  antisymmetric (AS) representations coincide:  $\Psi_{ij} = \epsilon_{ijk} \Psi_{k}$.  So, one can take a large-$N$ limit of QCD in either of two ways. With two index irreps, fermions are unsuppressed in the large-$N$ limit: they have $O(N^2)$ degree of freedom, just like gauge bosons, and unlike fermions in the fundamental representation. 
 The   exact equivalence in  the $N=\infty$ limit allows one to convert knowledge about supersymmetric theories into improved understanding of QCD or QCD-like theories.  
 For example, one immediate prediction is that the chiral condensate in ${\cal N}=1$ SYM and in QCD(AS) must agree up to $1/N$ corrections. The  condensate is calculable in ${\cal N}=1$ SYM \cite{Davies:2000nw}
 because it is an element of the chiral ring, so we learn something non-trivial about the non-supersymmetric QCD(AS) theory: 
$ \langle  \textstyle  {\rm tr}  \lambda \lambda \rangle^{\rm QCD(AS) } =
   \langle  \textstyle  {\rm tr}  \lambda \lambda  \rangle^{\rm SYM}  ( 1 + O(1/{N}) )$.

Another interesting aspect of  the equivalence  is that it sometimes has  {\it counterintuitive} consequences, which are conceptually  important in QFT.  
 A natural setting where this happens is the following.  Assume equivalence is valid. Then,   if either  the parent or daughter  theory has  a global symmetry which protects an observable, 
then in the other theory (which does not possess the same global symmetry)  the image observable  {\it behaves as if} it is protected by the higher symmetry of the companion 
\cite{Armoni:2007kd,Unsal:2007fb}. Why is this so?

Consider the following example of this kind of argument. In ${\cal N}=1$ SYM,  because of the  trace anomaly relation,  the gluon condensate determines the ground state energy, and as such, it is an order parameter for dynamical supersymmetry breaking: 
${\cal E}_{\rm vac} = \langle \Omega | T_{00} |\Omega\rangle =  \langle \Omega | 
\{Q_{\alpha}, {\bar Q}_{\dot \alpha}  \}  |\Omega\rangle  \sigma^{0}_ { \alpha \dot \alpha}
= \frac{1}{4} \langle \Omega |  T_{\mu \mu}   | \Omega \rangle   
= \frac{1}{4} \frac{\beta(g)}{g^3} \langle {\rm tr}\, F^2_{\mu \nu}\rangle$.
 This theory has non-zero Witten index, hence, supersymmetry is unbroken \cite{Witten:1982df}. This implies that the gluon condensate must vanish. 
As a result of equivalence, in non-supersymmetric  QCD(AS) the gluon condensate must {\it also} be zero, at leading order in the $1/N$ expansion, scaling as \cite{Unsal:2007fb}:
\begin{eqnarray}
 \langle  \textstyle \frac{1}{N} \tr F_{\mu \nu}^2  \rangle^{\rm QCD(AS) }
   = \frac{1}{N} \Lambda^4\ 
    \longrightarrow {\color{red}0} \qquad {\rm at}  \; N=\infty 
   \label{suppressed}
\end{eqnarray}   
 This is a very surprising  result!
 In confining $SU(N)$ gauge theories, the natural scaling of (properly normalized)  
 non-extensive observables is $O(N^0)$. 
For example,  the gluon condenstate in pure Yang-Mills theory  is parametrically 
 $\langle\textstyle\frac{1}{N}{\rm tr} F_{\mu \nu}^2\rangle^{\rm YM} \propto 
N^0\Lambda^4$, and one would expect to see this  ``natural" scaling in QCD(AS) as well.   Furthermore, QCD(AS) is non-supersymmetric, and hence one cannot use the tools of supersymmetry.  This means that, {\it even in a supersymmetric theory}, there must be a 
non-perturbative mechanism  for the vanishing  of the condensate, 
which does not rely on supersymmetry. Recall that both QCD(AS) and ${\cal N}=1$ SYM are vector-like theories and hence the fermion determinant is positive semi-definite. Also $ {\rm tr} F^2_{\mu \nu}$ is positive definite, in the Euclidean setting.  
This implies that the gluon condensate is the average of a positive observable  with respect to a positive measure \cite{Shifman:1978bx, Callan:1977gz, Schafer:1996wv}. 

\noindent
{\bf A puzzle:} The natural  QCD intuition is that the gluon condensate is positive-definite  in any confining  vector-like theory.   From the perspective of  supersymmetric QFT, the gluon condensate is zero in theories which do not break supersymmetry.  But ${\cal N}=1$ SYM is both supersymmetric and  vector-like, and it does not break supersymmetry.  This raises an interesting question:  is it possible to explain the vanishing of the gluon condensate without using the supersymmetry (SUSY) algebra, 
for example by using  semi-classical arguments?

This is the type of puzzling implication of large-$N$ orbifold/orientifold equivalence that is the most challenging.  Taking such challenges seriously may lead to a deeper understanding of 
QFT and path integrals, as we argue below. 
 The resolution of this puzzle, which we discuss in \S.\ref{HTA} below, challenges our present 
understanding of the semi-classical representation of the path integral.  Our resolution introduces 
 new concepts such as (contributing) complex saddles and  hidden topological angles.

We also mention that large-$N$ equivalences  also have useful applications to finite density QCD \cite{Cherman:2010jj, Cherman:2011mh}, and reveal a certain universality of the phase structure in different QCD-like theories \cite{Hanada:2011ju,Unsal:2007fb}.

\section{Large-$N$ Volume Independence}
{\bf Birth, Death and re-birth of large-$N$ reduced models:}  
(The subtitle  is borrowed from a recent talk by Gonzalez-Arroyo.) 
In 1982, Eguchi and Kawai  (EK)
came up with one of the most remarkable  ideas in large-$N$ gauge theory \cite{Eguchi:1982nm}. They proved that certain properties of $U(N)$ Yang-Mill theory, formulated on a periodic lattice with $\Gamma^4$-sites,  
are independent of the lattice size in the $N \rightarrow \infty$ limit, 
and can be reduced to a matrix model on $1^4$-site lattice,  
 {\it provided} center-symmetry in the matrix model is not spontaneously broken, and translation symmetry in the lattice field theory is unbroken \cite{Yaffe:1981vf}.  From the modern point of view, this is an orbifold equivalence. For example, starting with a unitary matrix model with $U(\Gamma^4N)$ gauge group, and orbifolding by a proper $({\mathbb Z}_\Gamma)^4$, one ends up with a four-dimensional quiver (or lattice) with $U(N)$ gauge theory on $\Gamma^4$ sites, \cite{Kovtun:2007py}. The orbifold  equivalence gives  a   mapping  of a matrix model to Yang-Mills theory, and if valid, can be useful both numerically and analytically.   
 
There was important progress in the early 1980s developing the EK idea
\cite{Gross:1982at, Parisi:1982gp,Das:1982ux}.   In the end, the original EK proposal fails due to broken (center) symmetry at weak coupling of the lattice theory (which is continuously connected to the continuum).  Two important modifications were introduced: (i) {\it quenched} (QEK)  \cite{Bhanot:1982sh}, and {\it twisted} (TEK) \cite{GonzalezArroyo:1982ub, GonzalezArroyo:1982hz}. Eventually it was understood that even these modifications were not enough to realize the full reduction, but a newer version of TEK works, see below. See   \cite{Bringoltz:2008av} for QEK, and  \cite{Azeyanagi:2007su,Teper:2006sp} for TEK; and  \cite{Lucini:2012gg}  for a review.
 
However, this subject is now not only  reborn,  it is also  flourishing, with both analytic and numerical progress, and useful 
spin-offs.  QCD(adj) and deformed-Yang-Mills \cite{Kovtun:2007py,Unsal:2008ch} and  a new version of TEK   \cite{GonzalezArroyo:2010ss} work. One important aspect of TEK that had already been known for some time was the relation to   non-commutative field theory \cite{Ambjorn:2000cs}. 
  Newer  and  related ideas connected to large-N reduction include the idea of adiabatic continuity, rigorous semi-classics, and  resurgence theory.   Each will be explored in subsequent sections.

  \vspace{0.3cm}
\noindent
{\bf Why does the original EK idea fail?} In an interesting way, the physical reason why the original EK reduction {\it does not work} predates the proposal itself.  And the reason why modern variants of the EK proposal {\it must work} also predates the proposal. 
These reasons can be seen without elaborate calculations, and can be explained almost without formulas. The main results are actually conceptual {\it realizations} rather than calculations.

Consider  compactifying Yang-Mills theory on $\R^4$ to $\R^3 \times S^1$, instead of to a four-torus $T^4$.  $S^1$ is a thermal circle, with circumference $\beta$ equal to the inverse temperature.  It is well-known that if we increase the temperature in pure Yang-Mills theory,  the theory moves  from the confined to the deconfined phase,  which is experimentally observed.  The associated symmetry is center-symmetry, ${\mathbb Z}_N$, under which the order parameter  Wilson line holonomy of the gauge field in the compact  $x_4$ direction, $\Omega({\bf x}) = {\rm P} e^{ i \int_{S^1} A_4({\bf x}, x_4)} $,  transforms non-trivially under ${\mathbb Z}_N: \tr\Omega({\bf x}) \rightarrow  h \tr\Omega({\bf x}), \; h^N=1$. 
 At small circle radius,  where the analysis is reliable due to weak coupling, Gross, Pisarski and Yaffe (GPY) calculated the one loop effective potential \cite{Gross:1980br}:
\begin{eqnarray} 
V[\Omega]={\color{red} (-1)} \times \frac{2}{\pi^2 \beta^4} \sum_{n=1}^{\infty} \frac{|\tr \,\Omega^n|^2}{n^4} <0  \qquad \Longrightarrow {\rm center\; instability}
\label{YM}
\end{eqnarray}
Clearly, the minus sign in front of the one-loop effective potential indicates the  center-instability of the deconfined phase, 
because the  potential is minimized for $\langle\frac{1}{N}\tr\Omega \rangle=1$.
This means that on $\R^3 \times S^1$, volume independence 
is achieved above a critical radius,   $\beta>\beta_c$,  but must fail  below a critical size $\beta<\beta_c$   \cite{Yaffe:1981vf,Kiskis:2003rd}. (This is sometimes called partial reduction.)
 This also means that even if the large-$N$ theory is compactified on 
$T^3 \times S^1_\beta$, where $\beta<\beta_c$, the equivalence must fail regardless of the radius of $T^3$. Nothing changes as one moves to a QCD-like theory. The message is:
\begin{quote}
  In  all QCD-like  theories, with  all representations of fermions,  and for all infinite Lie groups, Eguchi-Kawai reduction is destined to fail  in a thermal set-up at sufficiently    small thermal circle.  Thermal compactification  with $\beta<\beta_c$  is in direct conflict with large-$N$ reduction. 
\end{quote}

\subsection{Boundary Conditions as an Idea,  and Center Stability in QCD(adj)}
\noindent
{\bf Why it must  work?}
Despite the fact that thermal fluctuations are in conflict with the large-$N$ reduction idea, 
{\it quantum fluctuations} may lead to a different behavior.  We can motivate and illustrate the idea using supersymmetry, but in fact the eventual argument does not require SUSY.
Consider a gauge boson and a fermion in the adjoint representation of some gauge group $G$. Then, imposing periodic boundary conditions on fermions, boundary conditions which respect supersymmetry, one is guaranteed that a potential for the Wilson line is not induced to any loop order in perturbation theory.  
This vanishing hides a
simple secret.  In fact, the one-loop potential is 
\begin{eqnarray}
\qquad \qquad V[\Omega]={\color{red} (1-1)} \times \frac{2}{\pi^2 L^4} \sum_{n=1}^{\infty} \frac{|\tr \Omega^n|^2}{n^4} ={\color{red} 0} \qquad \Longrightarrow {\rm perturbatively \; marginal} \qquad 
\label{SYM}
\end{eqnarray}
The really important point   is not the cancellation itself, but the existence of the $+1$  part (which has {\it nothing} to do with supersymmetry).  It is just the consequence of periodic boundary conditions, $\lambda(L)=+ \lambda(0)$,  for fermions. This tells us that the quantum fluctuations (as  
opposed to thermal fluctuations)  of an adjoint fermion undoes the center-destabilizing effect of gauge fluctuations. This is the main idea underneath periodic boundary conditions. 
Taking one more step in this direction, i.e.,  with $n_f=2$  fermions, one ends up with center-stability which is the crucial necessary and sufficient condition for the validity of large-$N$ volume independence. 
\begin{eqnarray}
V[\Omega]={\color{red} (2-1)} \times \frac{2}{\pi^2 L^4} \sum_{n=1}^{\infty} \frac{|\tr \Omega^n|^2}{n^4} >0  \qquad \Longrightarrow {\rm center \; stability}
\label{QCDadj}
\end{eqnarray}  
This simple idea provides a solution of the Eguchi-Kawai proposal  in QCD(adj). Thanks to the orientifold equivalence between QCD(AS) and QCD(adj), this is also the resolution of the problem for QCD \cite{Kovtun:2007py}. In fact, even for $n_f=1$ (or ${\cal N}=1$ SYM), center-symmetry remains intact due to non-perturbatively induced neutral bion effects, as discussed in \S \ref{bions}. 

Lattice simulations with one-site or few-site  large-$N$ matrix models confirm this proposal \cite{Bringoltz:2009kb, Azeyanagi:2010ne,Hietanen:2009ex}, as does an earlier simulation with massive adjoint fermions \cite{Cossu:2009sq}, emulating $\R^3 \times S^1$ at $N=3$.
In the end, the resolution is a simple realization.  
We also note that  the potential Eq.\ref{QCDadj} was derived earlier in the context of gauge-Higgs unification in \cite{Hosotani:1988bm} in extra-dimensional model building, although this work did not discuss the implication of this potential for large-$N$ reduction or non-perturbative QCD(adj) dynamics.

\subsection{Perturbative Intuition Behind Volume Independence}  

Instead of providing a rigorous  background for volume independence, we would like to provide an intuitive perturbative explanation for why it works the way it does \cite{Unsal:2010qh}.  
The interesting point
 is the relation between the Kaluza-Klein spectrum of perturbative modes and its dependence on the  realization of center symmetry.

\begin{figure}[ht]
\begin{center}
\includegraphics[angle=0, width=1 \textwidth]{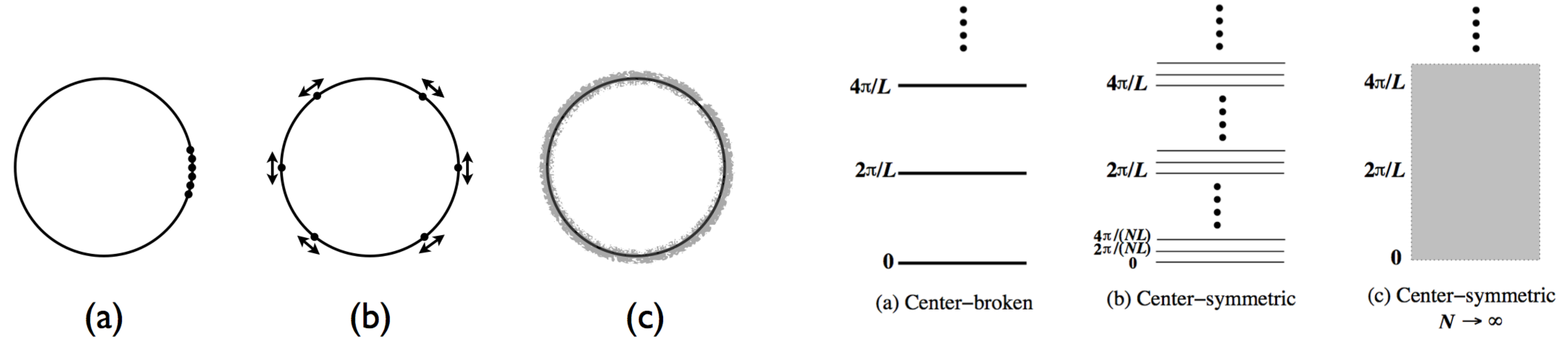}
\caption{Perturbative intuition of large-$N$ volume independence.  a) Center-broken holonomy, and  standard KK- spectrum;  b)  Center-symmetric holonomy  and finer 
 KK modes;
c) $N\rightarrow \infty$ limit of b). 
}
\label{vol-ind}
\end{center}
\end{figure}

If center-symmetry is broken, then the eigenvalues of the Wilson line clump together. 
Fig.~\ref{vol-ind}a  shows the standard form of Kaluza-Klein towers, in which discrete momenta 
are quantized in units of $\frac{2 \pi}{L}$, the usual text-book picture. 

If center symmetry is unbroken,  the eigenvalues  of the Wilson line  are evenly distributed around the unit circle. This results in a much finer ``KK-spectrum", due to  ``momentum-color intertwining".   The discrete momenta are quantized in units of  $\frac{2 \pi}{LN}$. See Fig.~\ref{vol-ind}b.

At fixed $L$, and as $N \rightarrow \infty$, nothing special takes place in the center-broken background or with the  standard KK spectrum.  But for the center-symmetric background,  the  spacing in the Kaluza-Klein spectrum approaches zero, and the spectrum becomes indistinguishable from the perturbative spectrum of the theory on $\R^4$! In other words, the scale $L$ disappears from perturbation theory as one takes the $N\rightarrow  \infty$ limit.  
One can interpret $LN\equiv L_{\rm eff}$ as the effective size of the compact dimension, and as one takes the $N\rightarrow  \infty$ limit, observables in the center-singlet sector (all local operators), and the spectrum of the theory, behave as if the theory is living on $\R^4$, despite the fact that the theory still lives on a space with size $L$. A similar  structure  appears in all toroidal compactifications on $\R^{d-k} \times (S^1)^k $, provided center symmetry is unbroken. With multiple dimensions compactified,  TEK realizes an even more efficient version of this color-momentum intertwining, due to non-commutativity of the background Wilson lines in different directions.  

\noindent
{\bf Scales,  two regimes, adiabatic continuity and universality:} 
 For center-symmetric confining theories on $\R^3 \times S^1$,  there are two different regimes:  
\begin{eqnarray}
&LN\Lambda \gg 1,   \;\; {\rm volume \; independent,  \; nonabelian \; confinement, \;incalculable}
\label{sa}
\\
&LN\Lambda \lesssim 1,  \;\;\;  {\rm volume \; dependent, \; \;\;\; abelian \;  confinement, \; \;\;\; \;\;\;   calculable} 
\label{sb}
\end{eqnarray}
Here, ``incalculable" means inaccessible with known analytical  techniques.  Since nature and numerical  lattice simulations know how to ``calculate" in this regime,  this may just mean that 
the  proper mathematical framework  and physical ideas are yet to be developed.   The first  regime Eq.\ref{sa} is the ``real thing", which we would like to understand. The second regime Eq.\ref{sb} is the  ``cartoon" of the real thing, but it is calculable.  These two regimes are very  often adiabatically connected, with no phase transition 
in between.  In cases where adiabatic continuity is believed to hold,  all non-perturbative observables  agree qualitatively, thus, the two regimes are in the same universality class.  But we cannot yet prove this in generality.

\subsection{A Puzzle and Emergent (Fermionic) Symmetry}  
A working large-$N$ volume independence  must have  a  dramatic spectral implication. 
  In QCD(adj), recall that periodic boundary conditions for fermions imply that, in the operator formalism,  we are calculating a {\it twisted (or graded)  partition function}, instead of the thermal partition function, namely, 
\begin{eqnarray}
\tilde Z (L) = \tr \left[  e^{-L  H} {\color{red} (-1)^F } \right] = Z_{\cal B} {\color{red} -}  Z_{\cal F} = \int dM [ \rho_{\cal B}(M) {\color{red} -} \rho_{\cal F}(M) ] e^{-LM}
\label{tpf}
\end{eqnarray}
This object is familiar from SUSY QFT, where it is  supersymmetric Witten index \cite{Witten:1982df}.  In non-supersymmetric theories,  its role has been understood only recently \cite{Unsal:2007fb, Unsal:2007vu, Unsal:2007jx,  Basar:2013sza}. It can be used to find the phases of a theory as a function of the spatial compactification radius (which now has no thermal interpretation). Unlike thermal compactification,  where the theory  must move to a deconfined phase at small circle-radius, 
in spatial compactification the quantum fluctuations measured as a function of radius may lead to very different small-$L$ behavior. In particular,  in QCD(adj), at infinite-$N$, there is no phase transition as a function of radius, i.e, $\frac{ d \tilde Z (L) }{dL}=0$.

Volume independence seems to be  puzzling in conjunction with the 
Hagedorn  spectrum of  hadronic states.   If a theory has Hagedorn growth \cite{Hagedorn:1965st}, with density of states 
$\rho(M) \sim \frac{1}{M} \left( \frac{T_H}{M}\right)^a e^{M/T_H} + \ldots$
where ellipsis are sub-leading power-law and exponential factors, 
 then the partition function must diverge above a critical temperature, $T_H\sim \Lambda$, parametrically related to the deconfinement temperature.
$T_H$ is the higher  limit of metastability, , see Fig.\ref{center-sym}, at which the meta-stable confined phase ceases to exist.  The actual  deconfinement temperature is slightly lower.
Regardless, the Hagedorn growth almost always demands the existence of a phase transition, 
while volume independence demands the opposite, the absence of any phase transition  \cite{Basar:2013sza}.  How can these two contradictory demands be reconciled? 

 The most reasonable resolution of this puzzle is an approximate spectral  
 degeneracy between the states in the bosonic and fermionic sectors, ${\cal B}$ and ${\cal F}$, similar to the $n_f=1$ SYM case.  Note that not only the leading Hagedorn growth, but all power law corrections and  all sub-leading exponentials must cancel  between the two factors. 
The resolution of such a high level of conspiracy  in the $n_f>1$ case seems to be feasible with  
the  emergence of a fermionic symmetry at large $N$, where there is no supersymmetry: namely,  
$\rho_{\cal B}(M) = \rho_{\cal F}(M)$, up to a possible $e^{M/(T_HN^p)}$ difference \cite{Basar:2013sza,Basar:2014jua}, $p \geq 1$. 
The potential conflict with the Coleman-Mandula theorem  is resolved because at $N=\infty$ the $S$-matrix is trivial, and hadrons are free.  

\section{Double-Trace Deformations:  Volume Independence vs. Adiabatic Continuity} 
\label{sec:adiabatic}
There is another solution to the volume independence proposal \cite{Unsal:2008ch}. 
So far, this approach does not seem to be as practically useful,  but conceptually it adds a new element at  large-$N$. Furthermore, at $LN \Lambda \lesssim 1$ it leads to semi-classical 
calculability in Yang-Mills and  QCD-like theories with one fermion flavor  \cite{Shifman:2008ja}, and a regime continuously connected  to large-$LN \Lambda$ and $\R^4$.

\noindent
{\bf Main idea \cite{Unsal:2008ch}:} Consider deforming, by the addition of a double-trace operator,  the Yang-Mills theory compactified on small $\R^3 \times S^1$: 
\begin{eqnarray}
&S^{\rm dYM}= S^{\rm YM} + S^{\rm dt} = \int_{\mathbb R^3 \times S^1}   \frac{1}{2 g^2} \tr F_{\mu \nu}^2 + 
  \int_{\mathbb R^3 \times S^1} \frac{1}{L^4} P[\Omega({\bf x})]  \cr \cr
  & P[\Omega]  = \sum_{n=1}^{\lfloor \frac{N}{2} \rfloor} a_n |\tr (\Omega^n)|^2, \qquad a_n > a_n^{\rm cr} >0
  \label{dtd}
\end{eqnarray}
 \begin{itemize}
\item The double-trace deformation is $O(N^2)$: it is as important as the action itself at large-$N$, and hence is {\bf not a small perturbation of the action.} 
\item In fact, it cannot be a small perturbation because it changes the phase structure  of the theory: it turns the deconfined phase into the confined phase. As such, one may at first guess that
the deformed theory has nothing to do with the original theory. This would be a natural point of objection for any sensible quantum field theorist.
\item But something striking happens.  The  effect of the deformation on observables  such as  the mass gap, spectra, string tension, topological susceptibility (neutral sector, center-singlet observables)
 is suppressed by $O(1/N^2)$. In other words, somewhat surprisingly, at large-$N$ {\bf  its effect on  observables is a  small perturbation of order  $O(1/N^2)$}.  For a proof of this statement using loop equations, see \cite{Unsal:2008ch}.
 
 \item In the volume dependent $LN \Lambda \lesssim 1$ regime, the deformed theory is analytically calculable, and this regime is adiabatically connected to the $LN \Lambda \gg  1$ regime, with no gauge-invariant order parameter that can distinguish the two regimes. This is the realization of adiabatic continuity.  
 \end{itemize}
In this sense, the double-trace deformation Eq.\ref{dtd} defies standard  intuition. It is a special structure.   Veneziano once coined for us the special status of these deformations: 
``{\it It is like a good samaritan, does the good deed, and disappears from the scene.}"  But why is this so? 

The reason is  $\delta S= {\color{black}+} \tr{\Omega}  \tr{\Omega}^{\dagger}$, is a {\bf positive  twisted  (center non-singlet) square operator}. The + sign  forces stability of the 
center-symmetry, for which  $\tr{\Omega}$   is an order parameter. In the Schwinger-Dyson (or loop)  equations, one of the operators enters in some connected correlator, but the other part is always a spectator. Factorization, which is guaranteed thanks to center-stability, is then invoked, and  results in: $ \langle \tr ({\rm stuff})   \tr{\Omega}^{\dagger}  \rangle = \langle \tr ({\rm stuff})  \rangle  \langle  \tr{\Omega}^{\dagger}  \rangle  + O(1/N^2)$. 
But  $ \langle  \tr{\Omega}^{\dagger}  \rangle $ is already zero because of  stabilization of the corresponding symmetry and the correction term  vanishes at $N=\infty$.  This is the magic of center-stabilizing double-trace deformation. This construction also generalizes to other symmetry stabilizing double-trace operators \cite{Cherman:2010jj}. Double-trace deformations have also been used in exploring {\it partial} center-symmetry breaking phases \cite{Ogilvie:2007tj, Myers:2007vc, Myers:2009df,Ogilvie:2012is}.

\subsection{Deformed Yang-Mills and adiabatic continuity} 
\label{adiabatic}
The dYM theory  in the  small-$S^1 \times \R^3$ regime  provides an example of semi-classsically calculable  confinement, which is adiabatically  connected  to pure Yang-Mills on $\R^4$, and strong coupling non-abelian confinement \cite{Unsal:2008ch,Shifman:2008ja}. 
 In this sense, it is one of the most useful arenas to study non-perturbative dynamics. 
To illustrate the basic idea, here we
discuss some non-perturbative aspects of $SU(2)$ gauge theories.  

At small $S^1$,  due to weak coupling and the deformation potential,  the  $SU(2)$ theory is Higgsed down to $U(1)$   by a center-symmetric Wilson line  
\begin{eqnarray}
 \Omega = \langle \Omega \rangle \times \left( \begin{array}{cc}
e^{i  A_4L/ 2  } & \cr
& e^{-i A_4L/2 }
\end{array} \right),   \qquad  {\rm with \; vev} \;\; 
\langle \Omega \rangle= \left( \begin{array}{cc}
e^{i  \frac{\pi}{2}} & \cr
& e^{-i \frac{\pi}{2}}
\end{array} \right). 
\label{hol}
\end{eqnarray}
where $A_4$ denotes fourth component of gauge field along Cartan subalgebra. 
 Since the center symmetry is the only global symmetry that can distinguish the small and large circle regimes, and it  is {\it unbroken}, the deformed theory exhibits {\it adiabatic continuity}.  
The Wilson line behaves as a compact adjoint Higgs field, in contrast with the
 Polyakov model in which the adjoint Higgs field is non-compact and algebra valued \cite{Polyakov:1976fu, Deligne:1999qp}.  This subtle difference in the topology of field space  has two significant consequences: first, instead of having just one type of monopole-instanton, there are two different types, ${\cal M}_{1}$ and ${\cal M}_{2}$. Second, there exists a 
 topological $\theta$ angle. 
One type of monopole-instanton is the regular  3d  one, and  the other is a twisted (affine)  one  \cite{Lee:1997vp, Kraan:1998sn}.These  defects carry two types of topological quantum numbers, magnetic and topological charge, $(Q_m, Q_T)$,  
give by ${\cal M}_{1} : (+1,  +\half)$ and ${\cal M}_{2} : (-1,  +\half)$. 
The monopole-instanton operators are given by 
\begin{eqnarray}
&&{\cal M}_{1} \sim  e^{ - S_0  + i \frac{ \theta }{2}}  e^{ -   \phi+i \sigma}, 
 \qquad 
{\cal M}_{2} \sim  e^{ - S_0 +  i  \frac{ \theta  }{2} } e^{ +   \phi - i \sigma }, \cr
&&\overline {\cal M}_{1} \sim  e^{ - S_0  - i \frac{ \theta }{2}}  e^{ -   \phi- i \sigma}, 
 \qquad 
\overline  {\cal M}_{2} \sim  e^{ - S_0 -  i  \frac{ \theta  }{2} } e^{ +   \phi + i \sigma },  \qquad  {\rm where} \;\; \;    \frac{4 \pi}{g^2}  (A_4L) \equiv  \phi \, .
\end{eqnarray}
Here  $\sigma$ denotes the dual photon field, defined through the abelian duality relation,  $  \epsilon_{ \mu\nu \lambda} \partial_{\lambda} \sigma  = {4 \pi L \over g^2 } F_{\mu\nu}$, and the form of the monopole-instanton amplitudes  account for long-range magnetic Coulomb interactions. Since fluctuations  $\phi$ around background holonomy field is gapped by deformation, we set it to zero in the  discussion of long-distance physics. 
 Note that the  action of the monopole-instanton  is $\frac{1}{N}$  of the 4d-instanton action, $S_0= \half \times S_I = \frac{8 \pi^2}{g^2N}$, with $N=2$ in the present example, and ${\cal I}_{\rm 4d}= {\cal M}_{1}{\cal M}_{2}=  e^{ - 2S_0 +  i \theta } $ is the 4d-instanton amplitude.

 \begin{itemize}
\item  At  $\theta=0$,  the theory on $\R^3 \times S^1$ realizes confinement and a mass gap due to the monopole-instanton mechanism \cite{Unsal:2008ch}, as in the  Polyakov  mechanism on $\R^3$ i.e., Debye screening  in a  magnetically charged plasma  \cite{Polyakov:1976fu}. The long-distance effective theory is 
\begin{eqnarray}
L^{\rm d} (\sigma) =&& \frac{1}{2L} \left( \frac{g}{4\pi} \right)^2  (\nabla  \sigma)^2  - 4 a  e^{-S_0}   \cos \sigma  
\label{lag1}
   \end{eqnarray}
   The mass gap (inverse Debye length)  is given by $m\sim L^{-1} e^{-S_0/2} = \Lambda(\Lambda L)^{5/6}$.
 
 \item Turning on the $\theta$ angle introduces a sign problem in the Euclidean description, and in the dual description in Eq.\ref{lag1}, alters  $ \cos \sigma \rightarrow     \cos \left( \textstyle \frac{\theta}{2} \right)  \cos \sigma  $. 
 For  $\theta=\pi$, 
the  monopole-instanton induced  gap dies off due to destructive topological interference, between ${\cal M}_{1}$ and $    \overline {\cal M}_{2}  $, and at leading order in semi-classics, the theory is gapless \cite{Unsal:2012zj}. 
To all orders in semi-classics the  theory must either be  gapless, or gapped and two-fold degenerate.  The latter choice is realized due to the magnetic bion mechanism, inducing a term  $e^{-2S_0} \cos(2 \sigma)$, originating from  the $[{\cal M}_{1}   \overline {\cal M}_{2} ] $ correlated 2-event \cite{Unsal:2012zj}.  The theory has two isolated vacua, and 
 exhibits  spontaneous CP-breaking.  This is indeed compatible with the theory on  $\R^4$, where the theory has a CP-symmetry at 
$\theta=\pi$,  which is  believed to be spontaneously broken  by the 
CP-odd condensate  $ \langle \tr\, (F_{\mu \nu} \widetilde F^{\mu \nu})  \rangle$.  This is again a manifestation of the adiabatic continuity idea. 

\item The fact that confinement is realized by a magnetic bion mechanism has a dramatic impact.   The theory has two kinds of domain walls (lines) L$_1$, L$_2$  for which  $|\int_{\rm L_i} d \sigma| =\pi$ crossing the line,  while for an external unit charge, the monodromy of the dual photon is $\oint d \sigma= 2\pi$. It is shown in \cite{Anber:2015kea} that the string is a composite, ${\rm L}_1 \overline {\rm  L}_2$ domain line, due to the $\cos 2 \sigma$ term.   
Consider a segment  of ${\rm L}_2$, and a well-separated  $q\bar q=  \overline {\rm L}_2 {\rm L}_1 $ pair parallel to L$_2$. Clearly,  $\overline {\rm L}_2 {\rm L}_1 $ can fuse into 
${\rm L}_2$, and reduce the energy of the system considerably. This forms  ${\rm L}_2  {\rm L}_1  {\rm L}_2$ line, where the joining points are the quark and anti-quark. Since the tension of ${\rm L}_1$  and ${\rm L}_2$ are exactly degenerate, separating quarks on the wall does not cost any energy, i.e, quarks become liberated on the wall.  Such a scenario was proposed by S.J. Rey and Witten in the mid 1990s \cite{Witten:1997ep} using the vacuum structure and ideas about confinement in supersymmetric theories.   The dYM theory and QCD(adj) (discussed in the next section)  are the first concrete  realization of this idea in a set-up  where the confining dynamics is analytically understood.  
\end{itemize}

\noindent The small circle limit of deformed YM  provides the  first  realization of the Polyakov mechanism  \cite{Polyakov:1976fu} and generalization thereof    in  a  locally 4d setting,  thirty  years  after the original idea \cite{Unsal:2008ch}!  As emphasized, dYM has intrinsically 4d aspects not present in the 3d Polyakov model.  There are a number of  interesting  recent works studying dYM, for example,   \cite{Thomas:2011ee, Thomas:2012ib, Thomas:2012tu, Zhitnitsky:2013hs, Li:2014lza},  and  \cite{Anber:2013xfa,Anber:2015wha,  Anber:2015bba}, and see  \cite{Vairinhos:2010ha, Vairinhos:2011gv} for  initial lattice studies.

\section{Magnetic and Neutral Bions in QCD(adj):  Confinement and Center-Stability}
\label{bions}
QCD(adj) compactified on $\R^3 \times S^1 $ does not break its center-symmetry when fermions are endowed with periodic boundary conditions, for any value of $n_f \geq 1$. 
In the small-circle regime, the theory has monopole-instantons, similar to dYM. 
To keep the analogy with the dYM example, we continue to work with $SU(2)$  theory. 
In contrast with the bosonic case,   
 each monopole-instanton operator has $2n_f$ fermion zero modes dictated by  Nye-Singer index theorem on $\R^3 \times S^1 $  \cite{Nye:2000eg, Poppitz:2008hr} (also see \cite{Poppitz:2009tw,
Poppitz:2009uq,  Bruckmann:2003ag, Misumi:2014raa,GarciaPerez:2009mg}).
  Thus, the monopole-instanton operators  become:
\begin{eqnarray}
{\cal M}_{1} \sim  e^{ - S_0 }  e^{ - \phi +i \sigma} \det_{I, J}  \lambda^I \lambda^J 
 \qquad 
{\cal M}_{2} \sim  e^{ - S_0  } e^{ +  \phi - i \sigma } \det_{I, J}  \lambda^I \lambda^J   \cr
\bar {\cal M}_{1} \sim  e^{ - S_0 }  e^{ -  \phi -i \sigma} \det_{I, J}   \bar \lambda^I \bar \lambda^J 
 \qquad 
\bar {\cal M}_{2} \sim  e^{ - S_0  } e^{ +   \phi + i \sigma }  
\det_{I, J}  \bar \lambda^I \bar \lambda^J 
\label{mon-op}
\end{eqnarray}
This implies that the monopole-instantons cannot induce a mass gap and confinement in QCD(adj) \cite{Unsal:2007jx}.  Note that the 4d instanton 
$I_{4d} \sim {\cal M}_{1} {\cal M}_{2}$  has $4n_f$ fermion zero modes.

At a superficial level, QCD(adj) on  $\R^3 \times S^1 $  resembles the Polyakov model coupled to massless Dirac fermions on $\R^3$. But the former confines, and the latter does not. 
To understand the difference,  first, recall the important result \cite{Affleck:1982as}. 

 \noindent 
{\bf {\color{black}  Affleck, Harvey, Witten [1982]}:} The Polyakov model  coupled to massless Dirac fermions on $\R^3$  remains gapless non-perturbatively, and does not confine.  
The work of AHW was viewed as the death of the Polyakov model in theories with fermions.
 Yet, in QCD(adj) on $\R^3 \times S^1$,  it has been shown that  the theory confines, and induces a mass gap for gauge fluctuations \cite{Unsal:2007jx}.  What is the major difference between the two? This is most easily explained in terms of global symmetries and the presence/absence of anomalies, and monopole-operators. 

On  4-manifold   $\R^3 \times S^1 $,   the theory has an anomalous chiral $U(1)_A$ symmetry, reduced down to  ${\mathbb Z}_{4n_f}$ due to instantons. The action of the ${\mathbb Z}_{4n_f}$ on the multi-fermion part of the monopole-operator 
Eq.\ref{mon-op} is negation, but since ${\mathbb Z}_{4n_f}$  is anomaly free, the invariance of the monopole operator demands one to shift $\sigma \rightarrow \sigma + \pi$, i.e. there is a ${\mathbb Z}_2$ topological  shift symmetry associated with monopole-operators which forbids $e^{i (2n+1) \sigma}$, but permits $e^{i 2n \sigma}$, $n\in {\mathbb Z}^{+} $.  Thus, the generation of a mass gap for gauge fluctuations is permitted \cite{Unsal:2007vu}.  In contrast, on a 3-manifold   $U(1)_A$ is non-anomalous,    and 
  the action of  $U(1)_A$  on the multi-fermion part of the monopole operator is continuous, say by $e^{i 4n_f \alpha}$.  The invariance of the monopole operator demands that  $\sigma \rightarrow \sigma - 4n_f \alpha$, and this continuous shift symmetry protects the gaplessness of the dual photon \cite{Affleck:1982as}. Relatedly,   ${\cal M}_{2}$ does not exist on 
  $\R^3$ \cite{Unsal:2007vu}.

  \begin{figure}[ht]
\begin{center}
\includegraphics[angle=0, width=0.8 \textwidth]{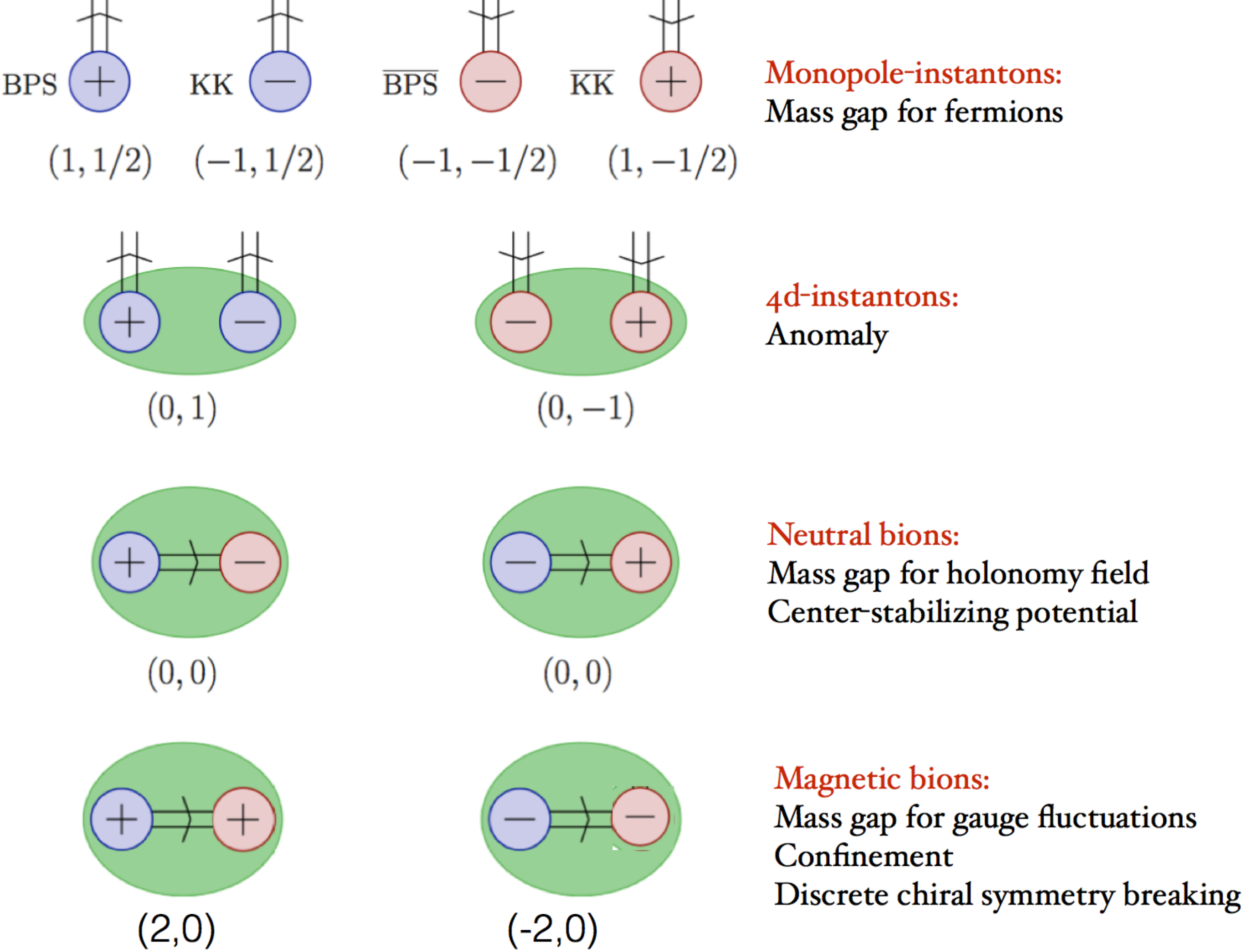}
\caption{Magnetic and topological charges 
$(Q_m, Q_T)$ of saddle-fields  in ${\cal N}=1$ SYM   and 
their role in non-perturbative dynamics. In QCD(adj), monopole-instanton has $2n_f$ fermion zero modes.
 }
\label{bionsetal}
\end{center}
\end{figure}

  \noindent
  On $\R^3 \times S^1$, at second order in the semi-classical expansion, there are two types of correlated two-events (or bion events):
\begin{eqnarray} 
&&{\rm magnetic \;  bions:}  \; {\cal B}_{12}=  [{\cal M}_1 \overline {\cal M}_2]
      \sim  e^{-2S_0} e^{+2 i \sigma}  ,   \qquad   \; {\cal B}_{21}=  [{\cal M}_2 \overline {\cal M}_1]
      \sim e^{-2S_0} e^{-2 i \sigma}, \cr    
&&{\rm  neutral \;  bions:  } \;\;\;\;  {\cal B}_{11}= [{\cal M}_1 \overline {\cal M}_1]
      \sim  e^{-2S_0 + \color{red} {i \pi}}e^{ - 2\phi} ,   \;  \; \; \;  {\cal B}_{22}= [{\cal M}_2 \overline {\cal M}_2]
      =      e^{-2S_0 + {\color{red}   i \pi}}  e^{+2 \phi}, \qquad \qquad
      \label{bions-amp}
      \end{eqnarray}
   Magnetic bions  \cite{Unsal:2007jx, Anber:2011de}  and neutral bions \cite{  Poppitz:2011wy,  Argyres:2012ka,  Argyres:2012vv}   are 
   correlated-two-events  with magnetic and topological charges $(\pm2, 0)$ and $(0,0)$.   The neutral bion is indistinguishable from the perturbative vacuum in the usual topological sense.  
   But below, we discuss the meaning of  the $e^{{i \pi}}$ factor, which arises as a {\it hidden topological angle},  which can distinguish the two saddles.

   The magnetic and neutral bion saddles are non-self-dual. If the monopole-size is $L$, the bion saddles have a characteristic size $L/g^2$, parametrically larger than the monopole size, but  parametrically smaller than the  inter-monopole separation.  The long distance effective theory is described by the proliferation of one- and two-events, and  is given by (for $n_f  > 1$ theories,  the holonomy field $\phi$ decouples from the IR-dynamics) 
 \begin{eqnarray}
L^{\rm d} (\sigma) =&& \textstyle{ \frac{1}{2L} \left( \frac{g^2}{4\pi} \right)^2 }  
[(\nabla  \phi)^2 + (\nabla  \sigma)^2 ]+  e^{-2S_0}  \left( -\cos {2\sigma}  -   e^{i\pi}  \cosh{2\phi} \right)  + (n_f-1)V (\phi)   \cr
&&   \textstyle{ + \bar \lambda^I  \sigma_i \partial_i  \lambda^I + 
e^{- S_0}  \cosh \phi  \cos  \sigma  (\det_{I, J}  \lambda^I \lambda^J + {\rm {c.c.}})  }
 \label{lag3}
   \end{eqnarray}
 Magnetic bions lead to a mass gap for gauge fluctuations and a finite string tension, i.e. confinement.   To our knowledge, this is the first analytic demonstration of confinement in a  locally 4d (non-supersymmetric) QCD-like gauge theory, and it is a relatively unanticipated mechanism. Neither monopole-instantons, as in Polyakov model \cite{Polyakov:1976fu}, nor monopole or dyon particles, as in Seiberg-Witten theory  \cite{Seiberg:1994rs}, is the origin of confinement. Instead, the mechanism is 
induced by non-selfdual bion saddles.    Furthermore, the bion-induced potential has two minima within the fundamental domain of $\sigma$, with $\langle e^{i \sigma} \rangle = \pm 1$ associated with discrete chiral  symmetry breaking and two vacua. 

Neutral bions generate a non-perturbative potential for the Wilson line, and generate a 
potential between the eigenvalues of Wilson lines, i.e, preferring center-stability. 
In the $n_f=1$ theory, since the perturbative contribution  to the Wilson line potential is absent, 
center stability is realized via the neutral bion mechanism. 

For $1 <n_f  <n_f^{*}$, where $n_f^*$  is the lower boundary of the conformal window,  the theory is expected to exhibit $SU(n_f) \rightarrow SO(n_f) $ non-abelian chiral symmetry breaking on 
$\R^4$.   At weak coupling, this symmetry is {\it not} spontaneously broken,  and the small-$S^1 \times \R^3$ regime realizes confinement with discrete chiral symmetry breaking, but without non-abelian chiral symmetry breaking \cite{Unsal:2007vu,Unsal:2007jx}.
 As one increases $L$, (i.e., at stronger coupling)  the monopole-induced term can lead to continuous chiral symmetry breaking, as in the Nambu-Jona-Lasinio model, but now realized as a zero temperature quantum phase transition. 
This happens at the boundary of the semi-classically justified long-distance effective  theory. Unfortunately, there is no known microscopically reliable method to explore this regime. Nonetheless,  it is encouraging  that the reliable effective theory  based on  monopole-instantons and bions  generates confinement, discrete chiral symmetry breaking and has the seed of non-abelian chiral symmetry breaking  on QCD-like theories on $S^1 \times \R^3$.   Models  emulating this physical set-up  also exhibit   
 non-abelian chiral symmetry breaking \cite{ Nishimura:2009me, Shuryak:2012aa, Liao:2012tw, Larsen:2015tso,Misumi:2014raa}. For other roles of  the bion saddles in diverse theories, 
 see  \cite{Misumi:2014bsa, Misumi:2014jua, Nitta:2015tua, Nitta:2014vpa}.

  \section{Toward Picard-Lefschetz Theory of Path Integration} 
  \label{picard}
Some aspects of the bion saddles in $\N=1$ SYM and QCD(adj) raises  interesting puzzles. 
For example, how can  bions (which are presumably  constructed as {\it approximate} solutions) be responsible for {\it exact} results such as the vanishing of vacuum energy or a condensate in  $\N=1$ SYM. Further, consider the following facts, which at first sight seem to suggest a possible incompatibility between SUSY and semi-classical analysis: 
  \begin{itemize}
  \item It is a widely known fact  that in supersymmetric theories, the ground state energy is  {\it positive-semidefinite non-perturbatively}, and zero to all orders in perturbation theory. 
\item It is an under-appreciated, but extremely important fact, that the contribution of {\it real} non-perturbative saddles (e.g. instantons)  to the ground state energy is  {\it  universally negative semi-definite} (in the absence of a topological $\theta$-angle or Berry phase)  \cite{Behtash:2015zha, Behtash:2015loa}. 

\end{itemize}
The standard textbook treatment of path integrals [see
the beautiful book by Coleman \cite{Coleman}, or  reviews on QCD \cite{Schafer:1996wv}],  presents a formalism in which only real saddles are accounted for.  
Does this mean that semi-classical analysis and supersymmetry are incompatible with each other? The answer is no, but the answer requires a significant revision of conventional semi-classics, broadening the perspective to explicitly include {\it complex saddles}.

\vspace{0.3cm}

\noindent
{\bf Euclidean  path integral in quantum mechanics:}
Assume one is given a Euclidean path integral  over real paths/fields.  Refs.\cite{Behtash:2015zha, Behtash:2015loa} propose that the semi-classical analysis of generic Euclidean path integrals necessarily requires complexification of the path space,  ${\cal P}$,
 \begin{eqnarray}\label{eq:comp}
 x(t) \in {\cal P} \longrightarrow z(t)= x(t) + i y(t) \in  {\cal P}^{\mathbb C} 
 \label{comp}
\end{eqnarray}
Here $ {\cal P}^{\mathbb C} $ denotes the complexified path space, and one should consider 
path integration over the middle-dimensional cycles $\Gamma$ for which  $\Re [ {\cal S} ( z(t))] \rightarrow \infty$.   
This means that in order to find the possible set of saddles that may contribute to the path integral,  instead of solving the {\it real} classical Euclidean equations of motion, one should solve the {\it holomorphic} classical Euclidean equation of motion. For example, in quantum mechanics of a particle in a potential $V(x)$,  
 to find saddles, one should solve the holomorphic Newton equations in the inverted potential:
 \begin{eqnarray}\label{eq:eom1}
\frac{d^2 z}{dt^2}= +\frac{\partial V}{\partial z},  
      \qquad {\rm instead \;  of }  \qquad   \frac{d^2 x}{dt^2}=+\frac{\partial V}{\partial x} 
\end{eqnarray}
where $V(z) = V_{\rm r}(x,y) +i V_{\rm i}(x,y)$ is holomorphic. 
The second one of these, the  textbook recipe, misses some of the most important contributing saddles. Note that the first equation in  Eq.\ref{eq:eom1} is a coupled set of equations, which generically do not decouple, for the real and imaginary parts of the path. This simple fact has interesting consequences \cite{Behtash:2015zha, Behtash:2015loa}.

The relation between holomorphization and Picard-Lefschetz theory is the following.  If we (boldly) translate standard knowledge of steepest descent cycles in finite dimensions \cite{AGZV2, Pham}
to infinite dimensions,  we find a complex version of gradient flow \cite{Witten:2010cx}. For the real space path integral, this amounts to solving the complex gradient flow equations:
\begin{eqnarray}
\frac{\partial z  (t, u)}{\partial u} 
  =  +  \frac{ \delta \bar {\cal S} }{ \delta \bar z}    
 =  + \left(  \frac{d^2 \bar z}{dt^2}  
            -  \frac{\partial\bar V}{\partial {\bar z}}   \right)     
      \, , \qquad
\label{PLW-3}
\end{eqnarray}
The holomorphic equations of motion are just the vanishing condition on the right hand side of the complex gradient flow (or Picard-Lefshetz) equations. The resulting solutions form a more complete basis for saddles which may contribute to path integral. 

The practical implications of these formal ideas have not yet been fully realized in QFT.
However, in the 
path integral formulation of quantum mechanical systems with bosonic and $n_f$ Grassmann fields (emulating the flavor degrees of freedom of QCD), it has been demonstrated that the counterpart of the magnetic bion is an exact {\it real} saddle  non-selfdual solution,  and the  counterpart of the neutral  bion is an exact  {\it complex} non-selfdual solution to the holomorphic Newton equations in the inverted potential \cite{Behtash:2015zha, Behtash:2015loa}.

\subsection{Hidden Topological Angles and Complex Saddles}
\label{HTA}

A {\it hidden topological angle} (HTA) is an  invariant angle associated with saddle points of the complexified path integral and their descent manifolds (Lefschetz thimbles)\cite{Behtash:2015kva, Behtash:2015kna}. 
 The HTA is  distinct from  theta-parameter  in the lagrangian.  
 But in a way similar to the usual $\theta$-parameter, which turns the instanton fugacity
 into a  complex fugacity, $e^{-S_I + i \theta}$, the HTA also plays a similar role in the path integral saddle space. For real saddles, the HTA is zero.  
The  HTA plays a crucial role in the dynamics, e.g, gluon condensate, vacuum energy,  and center stability, as  discussed below. It is a conceptually new ingredient in the more rigorous understanding of semi-classics.  

  Now, we can  explain the resolution of the problem that we set in \S\ref{largeN}:   the vanishing and anomalously small gluon condensate in, respectively,  ${\cal N}=1$ SYM and non-supersymmetric QCD(AS). This also provides the resolution of the puzzle stated at the beginning of \S\ref{picard}, concerning an apparent incompatibility of SUSY with semi-classical analysis.
To highlight the surprising aspect of this analysis,  we recall the folklore  that  
 the gluon condensate can only receive positive semi-definite contributions in a semi-classical expansion (when the topological theta angle is set to zero) in a vector-like theory,  see  
 \cite{Callan:1977gz, Schafer:1996wv, Shifman:1978bx}. 

Consider  ${\cal N}=1$ SYM on small ${\mathbb R}^3 \times S^1$ where it is semi-classically calculable.   At leading order in semi-classics, the gluon condensate is zero 
because in the  monopole-instanton background, the measure has two unpaired fermion zero modes, and the Grassmann integral over them gives zero. At second order, there are  non-vanishing  contributions. Magnetic bions contribute positively, but there is an extra  $e^{i\pi} $ phase associated with the neutral bion. As such, at second order in semi-classics, one has 
\begin{eqnarray}
\label{cancel}
L^4  \langle {\textstyle \frac{1}{N}} {\rm tr} F^2_{\mu \nu} \rangle  
   &=
    (n_{ {\cal B}_{ij}} +   e^{i  \pi} n_{ {\cal B}_{ii}})  = e^{-2S_0} +  e^{{\color{red} i \pi }} e^{-2S_0}  =  {\color{red} 0} \, . 
\end{eqnarray}
By the trace anomaly, this translates to a result for the vacuum energy density: 
\begin{eqnarray}
{\cal E}_{\rm vac}  = \frac{1}{4} \frac{\beta(g)}{g^3} \langle {\rm tr}\, F^2_{\mu \nu}\rangle \propto -e^{-2S_0} -  e^{{\color{red} i \pi }} e^{-2S_0}  =  {\color{red} 0} 
 \end{eqnarray} 
The  HTA, the angle $\pi$ in the $e^{i\pi}$ factor accompanying the neutral bion contribution, provides the resolution of both the puzzles mentioned in \S\ref{largeN} and \S\ref{picard}, which actually turn out to be the same problem due to the trace anomaly relation. 
The contribution of the real saddle to the vacuum energy is negative, but there also exists a complex saddle, whose action is complex due to the HTA, and its contribution to the vacuum energy is positive. Moreover, the real part of the actions of the real and complex saddles are the same, so the two contributions cancel. This is the mechanism by which the semi-classical expansion is consistent with the SUSY algebra. Both the complex saddle and the HTA are crucial for the argument. 
This example   also demonstrates that in  using  Lefschetz thimbles, for example, either in Euclidean   semi-classics  
or real time semi-classics (with sign problems) \cite{Tanizaki:2014xba, Cherman:2014sba}
or in lattice simulations 
\cite{Cristoforetti:2012su,Cristoforetti:2013wha,Fujii:2013sra,Aarts:2014nxa},  
all   thimbles whose Stokes multipliers are non-zero must be summed over. 
Numerical evidence for the  correctness of this perspective is also given in  \cite{Kanazawa:2014qma,Tanizaki:2015rda,Alexandru:2015xva, Hayata:2015lzj,Fujii:2015bua}.

In  QM models  with $n_f$ Grassmann fields,    it is possible to construct exactly the non-selfdual bion solutions. They are solutions to the second-order (non-BPS) classical equations, with finite  action, and the imaginary part of their action is always a multiple of $\pi$.   Furthermore, it has been shown that they are the  exact form of the (approximate) correlated two-instanton events, and can be identified with Lefschetz thimbles \cite{Behtash:2015kva, Behtash:2015kna}. Interestingly, if one restricts to real solutions it is not possible to find such exact saddle solutions, and  the standard practice in the past  was to construct approximate instanton/anti-instanton quasi-solutions. The situation is much improved now.

The relation between the instanton type BPS solutions and our exact non-selfdual solutions is also of importance and gives more insight into the full story. 
The position of a BPS instanton solution  is a  moduli, but with two such events, 
their separation is a quasi-zero mode, a parametrically small non-Gaussian mode, that needs to be integrated exactly to get the correct correlated two-event amplitude.   
 In the background of multi-instantons,  we consider decomposing the full space of fields  
  into 
 \begin{eqnarray}
{\cal J}^{\rm full} ={\cal J}^{\rm Gaussian} \times {\cal J}^{\rm zm} \times  {\cal J}^{\rm qzm}  ~.
\label{LT}
 \end{eqnarray} 
 where ${\cal J}^{\rm full}$ is the full thimble in field space, and ${\cal J}^{\rm Gaussian}$ are the Gaussian fluctuations of the fluctuation operator, $ {\cal J}^{\rm zm} $ and   $ {\cal J}^{\rm qzm} $  are exact and  quasi-zero mode directions. 
 As shown in \cite{Behtash:2015kna},  the relative angle between the magnetic and neutral bion  thimbles is
\begin{eqnarray} 
{\rm Arg} ( \mathcal{J}_{ {\cal B}_{ii}} )  
    = {\rm Arg} (\mathcal{J}_{ {\cal B}_{ij}} )  + 
     \pi\, . 
\end{eqnarray}
which is a ${\mathbb Z}_2^{\rm HTA}$-worth of hidden topological angle structure. 
This is the  origin of the vanishing of the gluon condensate and the vanishing vacuum energy  in the  semi-classical domain of ${\cal N}=1$ SYM.    The HTA  also explains the parametric smallness of the gluon condensate in non-supersymmetric QCD(AS)\cite{Behtash:2015kna}. Next, we discuss the role of the neutral bion and the HTA in center-symmetry realization.

\subsection{Calculability and the Deconfinement Phase Transition}
\label{calculability}
Studying deconfinement or the center-symmetry-changing phase transition in continuum  QCD-like theories  by reliable analytical methods proved to be impossible until recently. Perhaps the state of the art perspective  on the subject is still the one expressed by Gross-Pisarski-Yaffe (GPY) from 1981 \cite{Gross:1980br}: {\it ``It is hardly  surprising that  we cannot explore the transition as the temperature is lowered, from the unconfined to the confined phase using solely weak coupling techniques."}   
One possible way around this obstacle, by considering the theory on small  $S^3 \times S^1$, is found in \cite{Aharony:2003sx}.    In this context, since the theory is at finite volume, a genuine phase transition is only possible at  $N=\infty$, where the  thermodynamic limit  is achieved \cite{Aharony:2003sx}. 
 In  \cite{Poppitz:2012sw, Poppitz:2012nz}, a new idea resolving the GPY-problem  is presented on the thermodynamic   $\R^3 \times S^1$ setting,  so that it works  for arbitrary gauge group $G$. For related works, see \cite{Poppitz:2013zqa, 
 Anber:2014lba, Anber:2013doa, Anber:2012ig, Anber:2011gn,Shuryak:2013tka}.
\begin{figure}[ht]
\begin{center}
\includegraphics[angle=0, width=1.0\textwidth]{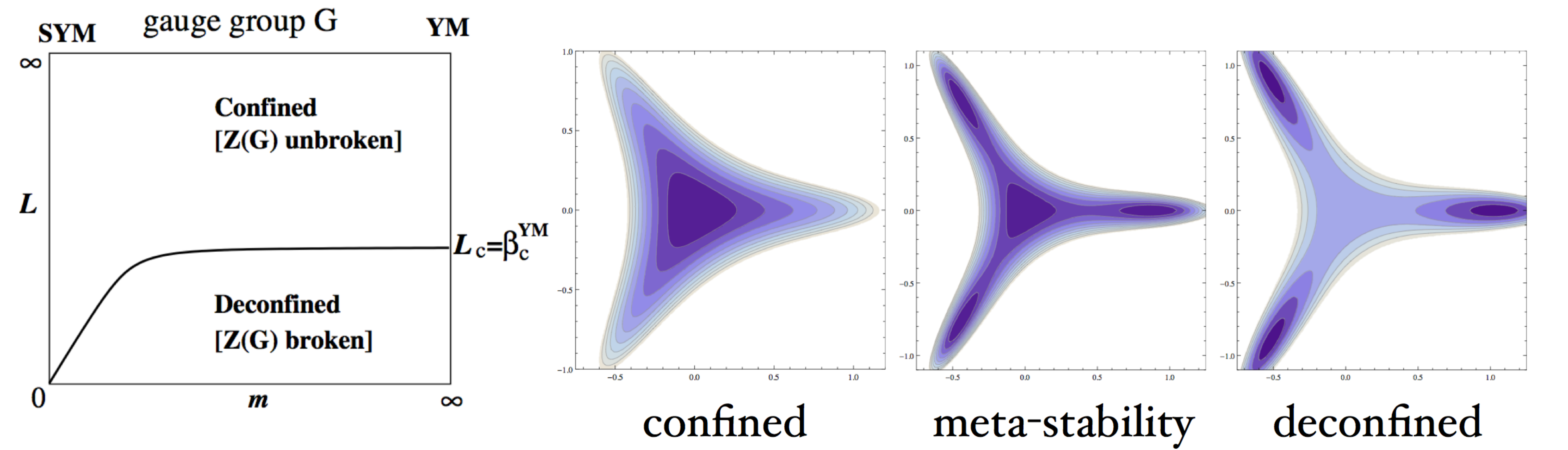}
\caption{(Left) Continuity of calculable and incalculable phase transitions. (Right) Contour-plot of the traced  Wilson line potential in three different (semi-classical) regimes, for $SU(3)$ gauge theory. 
}
\label{center-sym}
\end{center}
\end{figure}

 \noindent
{\bf The Main Idea \cite{Poppitz:2012sw, Poppitz:2012nz}:} The goal is to find a calculable phase transition and gain insight into the transition in pure Yang-Mills theory.  Consider ${\cal N}=1$ SYM, with a gauge boson and a fermion, and continuously connect this theory 
to pure Yang-Mills by turning on a mass term $m$ for the fermions.  
At $m=0$,   using the twisted partition function Eq.\ref{tpf}   on $\R^3 \times S^1$ (i.e., non-thermal periodic boundary conditions for fermions), this theory exhibits  analyticity   
as a function of $L$; hence, no phase transitions \cite{Witten:1982df, Davies:2000nw}.   In fact, Eq.\ref{tpf} is the Witten index, and is a constant, independent of $L$. 
 At $m=\infty$, the  theory reduces to pure YM. 
 Consider finite and small $(m, L)$. If one takes $m\rightarrow 0$,  $L=$ fixed, the theory lands on a  confining phase, and the center is unbroken due to the non-perturbative center-stabilizing potential induced  by the  neutral bion.  If one takes $m=$ fixed, 
$L \rightarrow  0$,  the theory lands on a center-broken (deconfinement) regime, 
due to the perturbative one-loop center-destabilizing  GPY-potential. This means that the two limits do not commute, and there must exist a calculable phase transition in between. Indeed there is, as shown in  Fig.~\ref{center-sym}(left), for any simple gauge group.  More precisely,    the deconfinement phase transition in pure Yang-Mills theory is continuously connected to a quantum (non-thermal)  phase transition in mass-deformed SYM theory. This transition can be studied in a controlled way. 
 
The surprising outcome of this idea is that  the mechanism governing the phase transition is universal, and valid for all simple groups. As an example, here we express the induced potential  in dimensionless variablles,  $\tilde m=m/\Lambda, \tilde L = L \Lambda$, just for the $SU(2)$ theory:
\begin{eqnarray}
\label{totalpotential2}
\tilde{V} &&= V_{\rm neutral \; bion} + V_{\rm magnetic  \; bion} + V_{\rm mon. \; inst.} + V_{\rm pert.} \\
&&
 \textstyle{ = { \color{red} + } \cosh 2 \phi - \cos 2 \sigma 
  { \color{red} - }  {1\over2}  {\tilde m   \over \tilde{L}^2 } \cos\sigma 
   \left( \cosh \phi - \frac{1}{3\log\tilde{L}^{-1}} 
            \phi \sinh \phi    \right) 
  { \color{red} -}  {1 \over 1728}  \left({\tilde{m}  \over  \tilde{L}^2}\right)^2 
         {1 \over  \log^3   \tilde{L}^{-1} } \; \phi^2 ~}. \nonumber
 \end{eqnarray}
Neutral bions  favor  center stability and generate repulsion among the eigenvalues of the Wilson line (forcing $\phi=\alpha=0$ in Eq.\ref{hol}.) Here, the crucial point is the HTA associated with the neutral bion Eq.\ref{bions-amp}. Were it not for the HTA,  the neutral bion  would lead to a center-destabilizing effect!  
At finite $m$, both  the perturbative one-loop  potential,  and 
the monopole-instantons  prefer center breaking, leading to an attraction among the eigenvalues of Wilson line. This contradicts some  folklore that  monopole-instantons favor confining Wilson line holonomy. 

The semi-classical phase transition is  driven by the competition between these three effects. Fig.\ref{center-sym}(right) shows  contour plots for  the potential for the trace of the $SU(3)$ gauge theory Wilson line  $\langle \tr (\Omega) \rangle $ 
in three regimes: (i) confined phase; (ii) phase between the limits of meta-stability (where  confined   or deconfined holonomy minima coexist, and one of these is the the  global minimum)  and (iii) deconfined phase.  
 All three regimes have been observed in lattice simulations \cite{Bringoltz:2005xx}. We remind the reader that this figure is not a model. There are no tuned parameters; it is the result of a controlled semi-classical computations.  
 
 This framework can also be used to investigate the effect of the 
topological  $\theta$-angle dependence on the phase transition and critical temperatures, see \cite{Poppitz:2012nz, Anber:2013sga}. The most interesting  results is that $T_c(\theta)$, the critical temperature,  is a multi-branched function, which has a minimum at  
 $\theta=\pi$.   These results  inspired various lattice studies, which exhibit remarkable matching to the theoretical predictions \cite{D'Elia:2012vv, Bonati:2013tt, D'Elia:2013eua, Bonati:2015sqt}. 
 The semi-classically calculable realization of center-symmetry also inspired related works in the strong coupling regime, see \cite{Liu:2015ufa, Liu:2015jsa, Larsen:2015vaa}, and lattice simulations \cite{Bergner:2014dua}.

\section{Resurgence, Renormalons and Neutral Bions}
\label{resurgence}

In this final section, we present a brief overview of the resurgence program  in non-trivial QFTs
 such as   bosonic non-linear  sigma models in two dimensions ($\mathbb {CP}^{N-1}$, Grassmannian, $O(N)$, principal chiral model (PCM)),  and  Yang-Mills  and one-flavor QCD-like theories  in four dimensions \cite{Dunne:2015eaa}.  The link to the previous sections is the fact that the neutral bion can be identified as the semi-classical realization of the infrared (IR) renormalons in these asymptotically free QFTs \cite{Argyres:2012ka, Argyres:2012vv, Dunne:2012ae, Dunne:2012zk}, and that the Lefschetz thimble construction naturally encodes resurgence.  These asymptotically free theories are incalculable on $\R^d$, ($d=2,4$), for reasons explained below, and are all believed to be gapped.  
We describe here the analysis for the two dimensional $\mathbb {CP}^{N-1}$ sigma model, 
and we comment later on  the similarities and differences for the other theories. In fact, the structure that emerges for all of these theories on $\R^{d-1} \times S^1$ exhibits a surprising level of universality. 

The deep puzzles concerning these theories, emanating from profound work in the  late 1970s and early 1980s \cite{'tHooft:1977am,Beneke:1998ui, Callan:1977gz, Schafer:1996wv},  are all very similar. 
Recently, progress has been made in  the non-perturbative understanding of these theories, by combining the idea of adiabatic continuity described in \S\ref{sec:adiabatic} above, and the mathematical formalism of resurgence.  Resurgence is a general mathematical formalism that unifies perturbative and non-perturbative expansions into a single unified framework in the form of a trans-series, such that the entire trans-series is internally self-consistent with respect to Borel summation and analytic continuation of the couplings, and naturally incorporates  Stokes phenomena \cite{Ecalle:1981,bh90,Costin:2009,Aniceto:2013fka,Dunne:2015eaa}. These ideas have also had a profound impact on the study of matrix models and string theory
\cite{Marino:2008ya,Pasquetti:2009jg,Aniceto:2011nu,Aniceto:2014hoa, Garoufalidis:2010ya}.  

It is well known that perturbation theory in almost all interesting QFTs  is a divergent asymptotic expansion, even after regularization and renormalization. A powerful method to extract physical meaning from such a divergent expansion is the Borel transform, and the Borel sum. The Borel transform maps the (typically factorially divergent) series to a new convergent series: $P(g^2) = \sum_{n=0}^\infty a_n\, g^{2n} \longrightarrow BP(t) := \sum_{n=0}^\infty a_n\, t^{n}/n!$,
and the Borel sum is the Laplace integral: $\P(g^2) =  \frac{1}{g^2} \int_0^\infty BP(t)\, e^{-t/g^2} dt $. 
Formally, $\P(g^2)$ assigns a value to $P(g^2)$, but 
in almost all interesting QM examples, and for the above QFTs, 
 $BP(t)$ has singularities at $t_i \in \R^+$ on the positive Borel axis, which means that $\P(g^2)$ as defined by the integral is \emph{ambiguous}, depending on how one deforms the contour around the singularity. In certain cases, such as the QM cubic oscillator or the Stark effect problem, there is such a singularity but it has a clear physical meaning in terms of the instability of a metastable state, so the ambiguity is easily resolved by the physical condition of causality. But in other theories, such as the symmetric double-well, or periodic potential, or in asymptotically free QFTs, the systems are stable,  and yet the ambiguous, non-perturbative imaginary contribution remains. In these cases, it implies that Borel resummed perturbation theory is incomplete on its own.
 Note that the pathological ambiguity has the non-perturbative form $\pm i e^{-t_i/g^2}$.  
 
In QM with instantons there is  another,  lesser known, pathology, coming not from perturbation theory but from  the non-perturbative "instanton-gas" expansion itself. The one-instanton sector is well-defined and unambiguous  \cite{Coleman}, but in the two-instanton sector, when one calculates the instanton-antiinstanton  $[\I \bar \I]$  amplitude, via a procedure that was referred to as the  BZJ-prescription in \cite{Argyres:2012ka,  Dunne:2012zk}, due to important works  by  Bogomolny \cite{Bogomolny:1980ur} and  Zinn-Justin \cite{ZinnJustin:1981dx},  one finds that  the $[\I \bar \I]$  amplitude is also multi-fold ambiguous. 
 This prescription was partly a black-box until recently, and did not always produced a sensible result. 
It is  now better understood using Lefschetz thimble  techniques and complexification of path integral \cite{Behtash:2015loa, Behtash:2015kva},  and again the net result is  that the two imaginary ambiguous non-perturbative terms cancel one another, restoring consistency of the full "trans-series" representation of the (real) physical quantity being computed:
\begin{equation}
\Im[ \P_\pm(g^2) + [\I\Ib]_\pm (g^2) ] = 0, 
\label{cancel}
\end{equation}
up to terms of higher order.    [This connection between different saddles is also related to an argument of Lipatov \cite{Lipatov:1976ny}, see also \cite{Balitsky:1985in}.]
  The ambiguities at leading order $e^{-t_1/g^2}$ cancel, and the sum of perturbative and non-perturbative sectors is meaningful, 
 ambiguity-free and accurate. In principle,  this is just the "tip-of-the-iceberg": e.g., it has now been shown in a class of QM models that these resurgent cancellations occur to all orders of the trans-series expansion \cite{Dunne:2013ada,Dunne:2014bca}. An obvious but deep question is: can this idea  work in QFT, in particular to cure the IR-renormalon problem of asymptotically free QFT? 

\noindent
{\bf 't Hooft's IR renormalon puzzle [1977]:} In  the asymptotically free QFTs mentioned above, there are  Borel singularities that are parametrically closer to the origin of the Borel $t$-plane, on $\mathbb R^{+}$,  than the  $[\I\Ib]_\pm$ singularity \cite{'tHooft:1977am, Beneke:1998ui}.  
The $[\I\Ib]_\pm$  singularity is 
related to the factorial combinatorial growth of the number of Feynman diagrams, while  the (more dominant)  IR renormalon singularity is related to divergences coming from phase space integration, at small internal momenta, smaller than  $\Lambda$.    't Hooft called  these singularities ``IR renormalons" in the hope that they would be  associated with a semiclassical saddle, such as an  instanton. 
For example, the $[\I\Ib]_\pm$ singularities are located at  $t=t_n= n (2S_I)$, but the IR-renormalon singularities are located at  $t = \tilde t_n  \sim  n (2S_I)/\beta_0 $, 
where $\beta_0$ is  the leading coefficient of the renormalization group beta function.  Due to this mis-matched factor of $\beta_0$, it appears that the cancellation mechanism at work in QM cannot possibly work in QFT on $\R^d$. However, there is another well-known problem in $\mathbb {CP}^{N-1}$, Yang-Mills etc., coming from the instanton-gas picture itself. There are instantons, but the so called ``dilute instanton gas approximation"  must be regulated, because the instanton size is a moduli parameter, whose integration leads to another IR divergence.  

{\bf The Main Idea \cite{Argyres:2012ka, Argyres:2012vv, Dunne:2012ae, Dunne:2012zk}:} The key step is to combine the concept of  adiabatic continuity \cite{Unsal:2007vu, Unsal:2007jx} discussed above, with  resurgence theory. 
Asymptotically free QFTs can be brought into a calculable semi-classical domain on  $\R^{d-1} \times S^1$, and with appropriate boundary conditions or deformation this can be continuously connected to the strongly coupled  domain of interest.
In the calculable  semi-classical domain, all  non-perturbative aspects of field theory  become tractable, and should not differ dramatically from the behavior on $\R^d$. 
 Once a small circle compactification respecting continuity is found, which is counter-part of center-symmetric background in gauge theory,   (note that high temperature limit of thermal compactification {\it never} achieves this),  something magical occurs. E.g., the 2d  instanton in the $\mathbb {CP}^{N-1}$ model, with action $S_I= \frac{4\pi}{g^2}$,   splits up into $N$  
kink-instantons with action $\frac{S_I}{N}= \frac{4\pi}{g^2N}$, which is now the minimal action semi-classical saddle  
\cite{Bruckmann:2007zh, Brendel:2009mp, Dunne:2012ae}.
At second order in semi-classical analysis, similar to Eq.\ref{bions-amp}, there are two types of bion: we refer to them as charged and neutral bions. The neutral bions  $\B_{ii} $ possess zero topological charge and zero ``magnetic" charge, and can thus mix with perturbation theory. In the zero flavor $n_f=0$ theory, the neutral bion amplitude is two-fold ambiguous, and Borel-resummed perturbation theory (which transmutes to the combinatorics in reduced twisted compactification   \cite{Dunne:2012ae, Anber:2014sda}
  is also two fold ambiguous. Remarkably, these two ambiguities cancel each other {\it exactly} \cite{ Dunne:2012ae, Dunne:2012zk}. 
Correspondingly, in the Borel plane the leading singularity is located at $t = \tilde t_1  \sim  (2S_I)/N $, which, since $\beta_0=N$ for  
$\mathbb {CP}^{N-1}$, is the same as the location of 't Hooft's elusive renormalon ambiguity  $t = \tilde t_1  \sim   (2S_I)/\beta_0 $: 
\begin{equation}
\Im[ \P_\pm(g^2) + [\B_{ii}]_\pm (g^2) ] = 0,  \qquad  \Im [\B_{ii}]_\pm \sim \pm  e^{- 2 \frac{4\pi}{g^2N}} \sim \pm \Lambda^2
\label{cancel-2}
\end{equation}
Here $\Lambda$ is the strong scale of the theory. 
Both the ambiguity and the jump  can be interpreted as Stokes phenomenon in  field space.

  
  \begin{figure}[ht]
\begin{center}
\includegraphics[angle=0, width=1.1\textwidth]{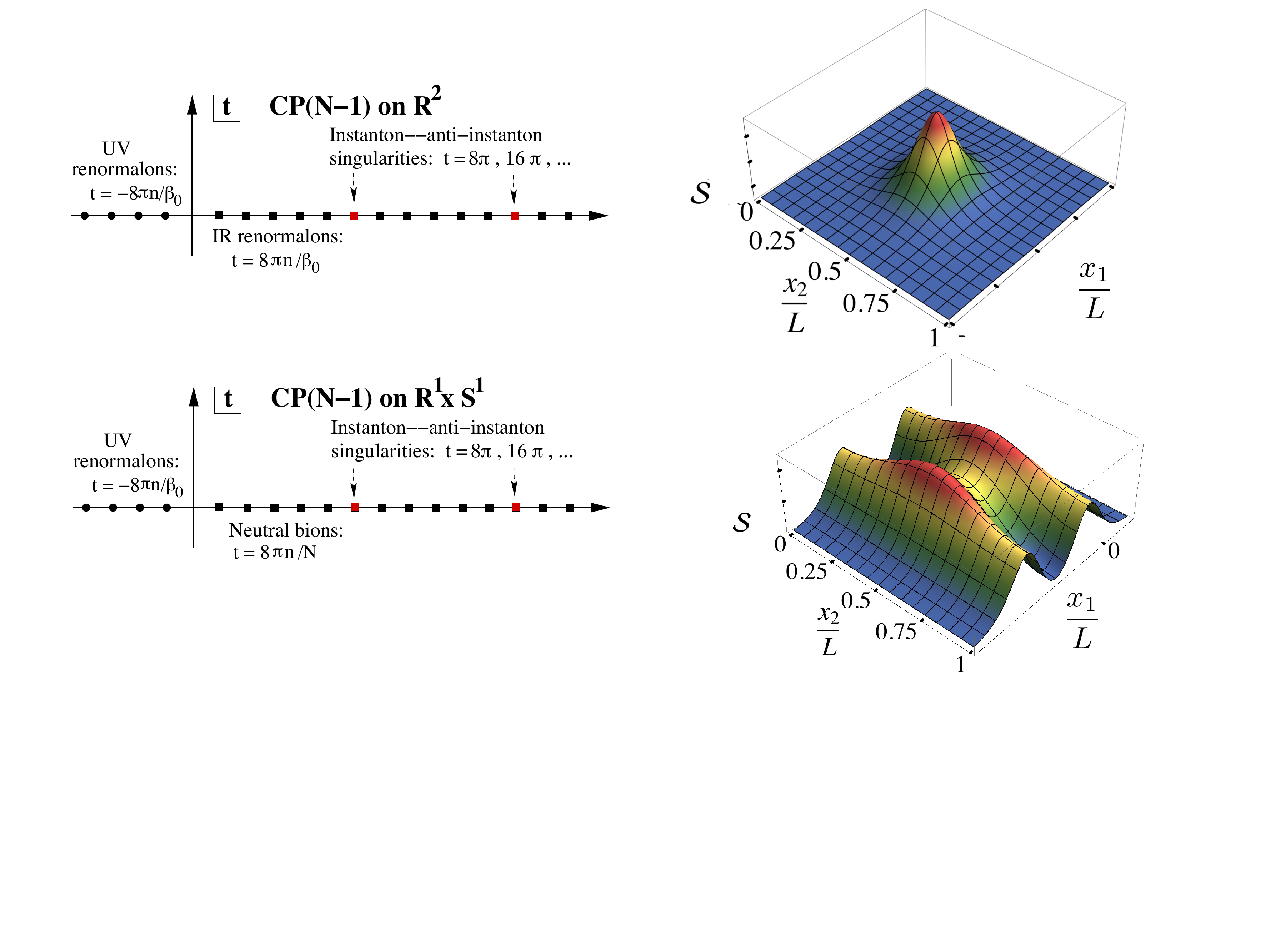}
\vspace{-4.3cm}
\caption{ (Left) Borel plane for $\mathbb {CP}^{N-1}$ on $\R^2$ versus the one in the weak coupling semi-classical domain on $\R^1 \times S^1$. The neutral bion singularities coincide with the IR renormalon ones. (Right) Splitting  of a 2d instanton  into two  1d kink-instantons as the size moduli is varied in the center-symmetric background on 
$\R^1 \times S^1$  for $N=2$ case. 
 }
\label{Borel}
\end{center}
\end{figure}

The mass gap in the small $\R^{1} \times S^1$ regime is also calculable, and  is given by  $m_g \sim (LN)^{-1} e^{- \frac{4\pi}{g^2N}} =\Lambda$. 
The gap is exactly what one expects on $\R^2$ based on large-$N$ and lattice simulations. It is induced at leading order due to  kink-instantons. This formula is valid for all $N$.

We conclude with some comments: 
\begin{itemize}
 \item The leading cancellation is the first step of a bigger structure. If these resurgent cancellations proceed ad infinitum, (as they are known to do in QM models), then in the weak-coupling semi-classical domain, our formalism provides a non-perturbative continuum definition of the QFT.

 \item In the $O(N \geq 4)$ vector model,  and in the principal chiral model, there are no instantons. So how can this resurgent cancellation idea work? In fact, these QFTs have finite action non-selfdual saddles. These saddles have negative fluctuation modes, producing imaginary factors, and play a role similar to the $[\cal I \bar \I]$-saddle \cite{Cherman:2013yfa, 
 Cherman:2014ofa,  Dunne:2015ywa}. Upon compactification, the 2d-saddles fractionalize into $N$ smaller action saddles.  Surprisingly, the saddle fractionalization structure in these theories on  $\R^{1} \times S^1$ is isomorphic to $SO(N)$  gauge theory on  $\R^{3} \times S^1$, following a pattern dictated by Lie algebra data, with multiplicities given by dual Kac labels \cite{Dunne:2015ywa}.

 \item In the calculable domain, puzzles  concerning "large-$N$ versus instantons"   \cite{Witten:1978bc}  simply  disappear , because the weight  of these saddles is governed by the 't Hooft coupling.

 \item
 By analogy with the operator product expansion, viewed as a trans-series combining non-perturbative condensates with perturbative fluctuations, one can use knowledge of the vanishing of certain condensates (e.g.,  due to some degree of SUSY) to predict novel cancellation mechanisms in perturbative sectors \cite{Dunne:2015eoa}.

\item In Yang-Mills (and other 4d gauge theories), the 
neutral bion singularities occur at $2S_I/N$ on small $S^1 \times \R^3$,   but the leading renormalon ambiguity is expected to occur at  $4S_I/\beta_0$ on $\R^4$. 
The crucial point (guaranteed by adiabatic continuity)  is that both singularities are at $\frac{1}{N}$ with respect to the instanton one. The   Borel flow of singularities   remain to be understood  \cite{Cherman:2014ofa}.

  \item 
We see tantalizing  evidence that observables in asymptotically free QFT are in fact resurgent. The goal would be to confirm that such a  resurgence property continues to hold for strongly coupled theories on $\R^d$ as well. 

\end{itemize}

\section*{DISCLOSURE STATEMENT}
The authors are not aware of any affiliations, memberships, funding, or financial holdings that
might be perceived as affecting the objectivity of this review. 

\section*{ACKNOWLEDGMENTS}
We  acknowledge useful discussions  with E. Poppitz, L. Yaffe, P. Argyres, M. Shifman, T. Sch\"afer,  A. Cherman, D. Dorigoni, G. Basar, T.  Sulejmanpasic, and M. Anber.  M.\"U.'s work  was  partially supported by the Center for Mathematical 
Sciences and Applications  (CMSA) at Harvard University. Our work was supported in part by the US Department of Energy grant   DE-SC0013036 (M.U.) and DE-SC0010339 (G.D.).

\end{document}